\documentclass[english,aps,pra,superscriptaddress,preprint]{revtex4}
\usepackage[T1]{fontenc}
\usepackage[latin9]{inputenc}
\usepackage{amsmath}
\usepackage{graphicx}
\usepackage{amssymb}
\usepackage{esint}

\makeatletter
\@ifundefined{textcolor}{}
{%
 \definecolor{BLACK}{gray}{0}
 \definecolor{WHITE}{gray}{1}
 \definecolor{RED}{rgb}{1,0,0}
 \definecolor{GREEN}{rgb}{0,1,0}
 \definecolor{BLUE}{rgb}{0,0,1}
 \definecolor{CYAN}{cmyk}{1,0,0,0}
 \definecolor{MAGENTA}{cmyk}{0,1,0,0}
 \definecolor{YELLOW}{cmyk}{0,0,1,0}
 }

\usepackage{dcolumn}\usepackage{bm}

\makeatother

\usepackage{babel}

\makeatother

\usepackage{babel}

\makeatother

\usepackage{babel}

\begin{document}

\title{Repulsive gravitational effect of a quantum wave packet and experimental
scheme with superfluid helium}

\author{Hongwei Xiong}
\email{xionghw@zjut.edu.cn}

\address{Wilczek Quantum Center, Zhejiang University of Technology, Hangzhou 310023, China}
\address{College of Science, Zhejiang University of Technology, Hangzhou 310023, China}

\begin{abstract}
We consider the gravitational effect of quantum wave packets when quantum mechanics, gravity, and thermodynamics are simultaneously considered. Under the assumption of a thermodynamic origin of gravity, we propose a general equation to describe the gravitational effect of quantum wave
packets. In the classical limit, this equation agrees with Newton's
law of gravitation. For quantum wave packets, however, it predicts
a repulsive gravitational effect. We propose an experimental scheme using superfluid helium to test this repulsive gravitational effect. Our studies show that, with present technology such as superconducting gravimetry and cold atom interferometry, tests of the repulsive gravitational effect for superfluid helium are within experimental reach.

\end{abstract}

\maketitle

\section{Introduction}

Although the unification of quantum mechanics and general relativity is elusive, considerable theoretical studies to reveal possible macroscopic quantum gravitational effect have been presented. A classic example is Hawking radiation \cite{HawkingRadiation}, predicted by combining general relativity, quantum mechanics, and thermodynamics. Although the quantum gravitational problem is far from solved, various approaches are being used to find evidence of quantum gravitational effects in high-energy scattering experiments and astronomical observations \cite{Ame,Jacob}. Many novel ideas are also proposed to test the quantum gravitational effects, e.g., probing the Planck-scale physics with a mechanical oscillator \cite{Planck}, the search for dark energy using atom interferometry \cite{Perl}, and the measurement of homological noise \cite{Hogan}. At present, no evidence of quantum gravitational effects have been observed in experiments on Earth.

In this study, we consider the gravitational effects for a quantum wave
packet based on the thermodynamic origin of gravity \cite{Jacobson}. By considering the fact that the direction of force can be naturally interpreted in thermodynamics, we find some justification for the attractive characteristics of Newton's law of gravitation. With the same considerations, we discuss a possible repulsive gravitational effect for quantum wave packets.
It is clear that, without a well-defined solution to the quantum gravitational
problem at the Planck length, this phenomenological theory requires
experimental testing. Fortunately, our studies show that current
techniques for measuring the gravitational force such as superconducting gravimetry and cold atom interferometry are able to test repulsive gravitational effects in superfluid helium.

The paper is organized as follows. In Section 2, we provide an explanation of the attractive gravitational force in Newton's law of gravitation. Section 3 is devoted to the consideration of the gravitational effects for a quantum wave packet when both quantum mechanics and thermodynamics are considered. We present a general equation
for the gravitational force for a quantum wave packet. In Section 4, using the general equation presented in the third section, we propose an experimental scheme to test the repulsive gravitational effect for superfluid helium. A brief summary and a discussion are presented in the final section.

\section{Thermodynamic understanding of the direction of gravitational force in Newton's law of gravitation}

It is well known that the law of gravity closely resembles the laws of
thermodynamics and hydrodynamics \cite{Bekenstein,Bardeen,Hawking,Davies,Unruh,Pand}.
This has led to intensive studies \cite{add,add2} into the thermodynamic origin of gravitation \cite{Jacobson}. When the thermodynamics of a system are considered, the direction of the force (e.g., pressure) can be determined from the thermodynamic properties of the system. This leads to a question about whether there is a physical mechanism determining the direction of the gravitational force, if the thermodynamic
origin of gravitation is assumed. In this section, we provide possible answers to this question.

Recently, the thermodynamic understanding of gravitation has been significantly advanced by Verlinde's work \cite{Verlinde}, in which the change of entropy $S$ after a displacement $x$ is given by the following formula:

\begin{equation}
S=2\pi k_{B}\frac{mc}{\hbar}x.\label{entropy}
\end{equation}
Here, $m$ is the mass of a fundamental particle, $c$ is the speed of light, and $k_{B}$ is the Boltzmann constant. In the original work
by Verlinde, this postulation (motivated by Bekenstein's work \cite{Bekenstein} regarding black holes and entropy) plays a key role in deriving Newton's law of gravitation. The proportional relation between $S$ and $x$ can be partially explained by an entropy increase with information loss. For a particle's motion in the vacuum background, when the particle arrives at a location after a displacement $x$, the information regarding its path is lost, which leads to an entropy increase when the vacuum background is also included.

Although there is significant controversy regarding the meaning and validity of this thermodynamic formula, it deserves further study. Considering
a particle with acceleration $\mathbf{a}$, we have
\begin{equation}
x_{j}=\frac{a_{j}t^{2}}{2}.\label{x}
\end{equation}
Here, $j = 1, 2, 3$. In this paper, all bold symbols represent vectors.
From Eq. (\ref{entropy}), we have

\begin{equation}
dS=\frac{2\pi k_{B}mc}{\hbar}\sqrt{\sum_{j}\left(a_{j}t\right)^{2}}dt.\label{ds}
\end{equation}
 In addition, from $E = \sqrt{m^{2}c^{4}+p^{2}c^{2}}$, using the nonrelativistic approximation, we have
\begin{equation}
dE=m\sum_{j}a_{j}^{2}tdt.\label{de}
\end{equation}
Using the fundamental thermodynamic relation $dE = T_{V}dS$, we have

\begin{equation}
T_{V}=\frac{\hbar}{2\pi k_{B}c}\frac{\sum_{j}a_{j}^{2}}{\sqrt{\sum_{j}a_{j}^{2}}}=\frac{\hbar\left\vert \mathbf{a}\right\vert }{2\pi k_{B}c}.\label{Unruh}
\end{equation}

It is clear that this temperature $T_{V}$ is different from the ordinary
temperature for an ensemble of particles in thermal equilibrium. In
this acceleration process, there is no entropy increase for the particle itself. In the acceleration process of the particle, the only possibility for entropy increase comes from the vacuum, if Eq. (\ref{entropy}) is correct. Thus, $T_{V}$ refers to the vacuum temperature at the location of the particle. This shows that the acceleration of a particle will induce vacuum excitations, and thus, lead to finite vacuum temperature at the location of the particle. Because there is a decay of these vacuum excitations in the propagation process, $T_{V}$ should be a spatially dependent temperature field distribution with a maximum value
given by $\hbar\left\vert \mathbf{a}\right\vert /2\pi k_{B}c$. As we do not have exact knowledge at the Planck length, it is not within the scope of the present work to provide the exact spatial scale of the decay. However, at least in the present work, the maximum value of $T_{V}$ is the physical quantity we need.

It is noteworthy that the above equation is the same as the Unruh temperature \cite{Unruh}.
This indicates the self-consistency of the above derivation. In Verlinde's work, both Eqs. (\ref{entropy}) and (\ref{Unruh}) are used to obtain Newton's law of gravitation. Here, we show the possibility of further simplification because Eq. (\ref{Unruh}) can be derived from Eq. (\ref{entropy}).

\begin{figure}
\centering \includegraphics[width=0.8\textwidth]{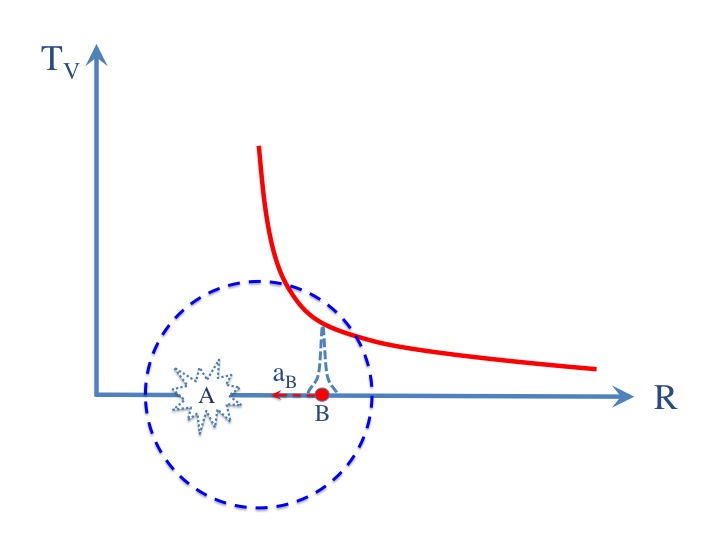} \caption{The relation between the vacuum temperature field distribution and acceleration. The red line shows the vacuum temperature field distribution for a system denoted by A, e.g., a celestial body. The red sphere represents an initially at rest classical particle at a location B. The acceleration $\mathbf{a}_{B}$
of particle B induces a vacuum temperature
field distribution (indicated by the dashed cyan line) with a peak value given by Eq. (\ref{eq:U2}). The local thermal equilibrium requests that this peak value is equal to the temperature of the vacuum temperature field (red line) at location B. This shows the physical mechanism that causes particle B to accelerate in the presence of a finite vacuum temperature field. For the whole system (surrounded by the dashed blue circle), the application of the free energy in the canonical ensemble determines the direction of the acceleration. With this consideration, the direction of the acceleration is determined by the gradient of the vacuum temperature field distribution, indicated by the dashed red line.\label{fig:acceleration}}
\end{figure}

The meaning of Eq. (\ref{Unruh}) is further explained by Fig. \ref{fig:acceleration}.
For a particle at location B with acceleration $\mathbf{a}_{B}$ (shown
by the dashed red arrow), the coupling between the particle and the
vacuum modes establishes the vacuum temperature field distribution (shown
by the dashed cyan line) with a peak value of $T_{V}\left(\mathbf{a}_{B}\right) = \hbar\left\vert \mathbf{a}_{B}\right\vert /2\pi k_{B}c$.
The strong coupling between the particle and the vacuum
modes leads to a ``dressed'' state that includes the local vacuum
excitations and the particle itself. If the particle has no size,
roughly speaking, the width of the local vacuum excitations is of
the order of the Planck length (using the same treatment for the
derivation of Newton's law of gravitation used in the latter part of
this section).

Eq. (\ref{Unruh}) shows the vacuum temperature field due to an accelerating
particle. It is natural to consider the opposite problem: What is the
acceleration of a particle in the presence of a finite vacuum temperature
field distribution? In Fig. \ref{fig:acceleration}, we assume that there is a vacuum temperature
field distribution (shown by the red line) that is due to a system
denoted by A, e.g., a celestial body. At location B, there is a particle (denoted by a red sphere).
To establish local thermal equilibrium, the red sphere will accelerate
such that the peak vacuum temperature of the dressed state is equal
to the temperature of the vacuum temperature field at location
B, ${\it {i.e.}}$ $T_{V}\left(\mathbf{a}_{B}\right) = T_{V}\left(B\right)$.
In this situation, we have
\begin{equation}
\left\vert \mathbf{a}_{B}\right\vert = \frac{2\pi k_{B}cT_{V}\left(B\right)}{\hbar}.\label{eq:U2}
\end{equation}

However, the acceleration is a vector, whereas the temperature is a
scalar. Therefore, the above formula does not define the direction
of the acceleration. For the whole system (including system $A$
and particle $B$ surrounded by the dashed blue circle), it is clear
that the particle number is conserved while there is energy exchange, which is
due to the propagation of the vacuum excitations. In this case, the
canonical ensemble should be used in thermodynamic studies of
the whole system. Therefore, we consider the free energy of the whole
system, which is defined as $F = U - T_{V}S$. Here, $U$ is the overall
energy of the system, which is a conserved quantity. Obviously, the
whole system is a non-equilibrium system. In thermodynamics, it is well established that system evolution has a tendency to decrease
the free energy in the most effective way. Under this consideration,
we use the following formula to determine the magnitude and direction
of the acceleration in the presence of a vacuum temperature field:
\begin{equation}
\mathbf{a}\left(\mathbf{R}\right) = \frac{2\pi k_{B}cT_{V}\left(\mathbf{R}\right)}{\hbar}\frac{\nabla_{\mathbf{R}}T_{V}\left(\mathbf{R}\right)}{\left\vert \nabla_{\mathbf{R}}T_{V}\left(\mathbf{R}\right)\right\vert }.\label{Unruhvector}
\end{equation}
 Here, $\mathbf{R}$ denotes a three-dimensional spatial vector, and $\nabla_{\mathbf{R}}$ denotes the vector differential operator. In Fig. \ref{fig:acceleration}, for a particle at location B, the
direction of the acceleration is based on the consideration
of the free energy. We will show that the above equation
provides an explanation of why the gravitational force is attractive between two spatially separated objects.


\begin{figure}
\centering \includegraphics[width=0.8\textwidth]{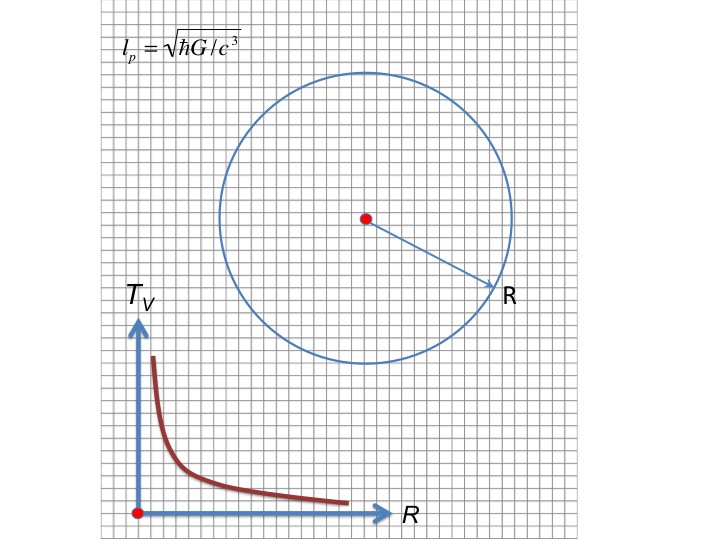} \caption{The grid shows the Planck length due to the quantum
gravitational effect. The red sphere shows a particle that leads to a vacuum
temperature field distribution through coupling with the vacuum.
The vacuum temperature field distribution is also shown in this figure.
\label{fig:Newtonlaw}}
\end{figure}

Now, we consider Newton's law of gravitation under the assumption of the
thermodynamic origin of gravitation. In Fig. \ref{fig:Newtonlaw},
we consider a space with a Planck length $l_{p} \equiv \sqrt{\hbar G/c^{3}}$, which shows the structure of the phase space due to the quantum gravitational effect.
Here, $G$ is the gravitational constant. Although the microscopic
mechanisms of $l_{p}$ are not completely clear, it is not unreasonable
to assume the existence of $l_{p}$ in the phase space. We assume
that the energy of a particle (indicated by a red sphere in Fig. \ref{fig:Newtonlaw})
is $\varepsilon$. We further assume that the number of degrees of freedom is $i$ for the elementary cells. From the
local thermal equilibrium, at the location of the particle, the temperature
of the vacuum is given by
\begin{equation}
\frac{i}{2}k_{B}T_{V}\left(R<l_p\right) = \gamma\varepsilon.
\end{equation}
Here, $\gamma$ denotes a dimensionless coupling strength between
matter and space. From ordinary statistical mechanics, $\mbox{\ensuremath{\gamma}}$
should be of the order of $1$. We have
\begin{equation}
T_{V}\left(R<l_p\right) = \frac{2\gamma\varepsilon}{ik_{B}}.\label{eq:Tvr0}
\end{equation}

When $R\rightarrow\infty$, $T_{V} = 0$. Thus, one expects the vacuum temperature
field distribution shown in Fig. \ref{fig:Newtonlaw}. For another
particle with mass $m$ in this vacuum temperature field distribution, from
Eq. (\ref{Unruhvector}), the acceleration field distribution is then
\begin{equation}
\mathbf{a}=-\frac{2\pi k_{B}cT_{V}}{\hbar}{\bf \mathbf{e}}_{R}.\label{eq:ar}
\end{equation}
 Here, the radial unit vector $\mathbf{e}_{R} \equiv {\bf \mathbf{R}/\left|\mathbf{R}\right|}$. The negative sign in the above equation derives from both the gradient of $T_V({\bf R})$ in the relation between $\bf a$ and $T_V({\bf R})$ and the relation in which a gradual decrease in $R$ results in an increase in $T_V({\bf R})$.  To obtain the above expression, the spherical symmetry of the system
for $R\gg l_{p}$ is also used. This equation explains the attractive gravitational
force between two classical objects. It is worth pointing out that in Newton's law of gravitation,
the attractive gravitational force is based on observations, rather than microscopic mechanisms. One of the merits of thermodynamics is that even if we do not know the exact collision properties (such as the scattering length between atoms), the macroscopic forces (such as
pressure) can be derived. When the thermodynamic origin of gravity
is assumed, we should expect the correct result for the direction of the gravitational force.

The continuous property of the gravitational force adds a further constraint
on the asymptotic behavior of the vacuum temperature field distribution for
$R\gg l_{p}$. For this reason, we have
\begin{equation}
T_{V}\left(\mathbf{R}\right) = \frac{\eta}{R^{2}}.
\end{equation}
 Note that the above expression holds for $R\gg l_{p}$; therefore, the
spherical symmetry approximation can be used.

Combined with Eq. (\ref{eq:Tvr0}), we have $\eta/l_{p}^{2} = 2\beta\gamma\varepsilon/ik_{B}$.
We have introduced a factor $\beta$ to provide a more rigorous derivation. Although we do not know the exact value of $\beta$, the asymptotic behavior of $T_V$ indicates
that $\beta$ is of the order of $1$. In this situation, we have
\begin{equation}
T_{V}\left(\mathbf{R}\right)=\frac{2\beta\gamma Mc^{2}l_{p}^{2}}{ik_{B}}\frac{1}{R^{2}}.
\end{equation}
 To get the above expression, we used $\varepsilon = Mc^{2}$.
Using Eq. (\ref{eq:ar}), we have
\begin{equation}
\mathbf{a}=-\frac{4\pi\beta\gamma c^{3}l_{p}^{2}}{i\hbar}\frac{M}{R^{2}}{\bf \mathbf{e}}_{R}.\label{eq:ar-1}
\end{equation}
 Because the gravitational constant $G$ is a physical constant obtained from observations, the factor
$4\pi\beta\gamma/i$ can be absorbed into it. Finally, we obtain the standard
result of Newton's gravitation law
\begin{equation}
\mathbf{a}=-\frac{GM}{R^{2}}{\bf \mathbf{e}}_{R}.
\end{equation}

We consider above-described the situation for a particle whose size is of the
order of the Planck length. If the size of the particle is larger
than the Planck length, using the Gauss's flux theorem for $R$
much larger than the size of the particle, the above result
still holds. Without an exact theory at the Planck scale, it is obvious
that the above derivation of Newton's law of gravitation is not
a rigorous derivation from first principles. Nevertheless, we have some evidence of the physical mechanism of the attractive gravitational
force in Newton's law of gravitation. In the next section, this idea
will be applied to study the gravitational force of quantum wave packets.

For an assembly of classical fundamental particles (here ``classical\textquotedblright{}\ means
that the quantum wave packet effect is negligible), we assume that the
vacuum temperature field distribution due to the $i$th particle is $T_{Vi}\left(\mathbf{R}\right)$.
Because there is no quantum interference effect between different
classical particles, the force
on an object with mass $m$ is

\begin{equation}
\frac{\mathbf{F}\left(\mathbf{R}\right)}{m}=\mathbf{a}\left(\mathbf{R}\right)=\frac{2\pi k_{B}c}{\hbar}\sum_{i}\frac{T_{Vi}\left(\mathbf{R}\right)\mathbf{\nabla}_{\mathbf{R}}T_{Vi}\left(\mathbf{R}\right)}{\left\vert \mathbf{\nabla}_{\mathbf{R}}T_{Vi}\left(\mathbf{R}\right)\right\vert }.\label{classicalacc}
\end{equation}
 The above expression is based on the assumption of the linear superposition
of gravitational forces ${\bf {\bf \mathbf{F}} = \Sigma}_{i}\mathbf{F}_{i}$.
In Newton's law of gravitation for an assembly of classical particles,
this implicit assumption is also used. We stress here that, rigorously
speaking, the summation in the above expression is about all fundamental
particles.

\section{Abnormal gravitational effect for a quantum wave packet}

In the thermodynamic origin of gravity, for an object with mass $M$,
it establishes a vacuum temperature field $T_{V}\left(\mathbf{R}\right)\sim M/\left\vert \mathbf{R}\right\vert ^{2}$.
Using the formula $\mathbf{a} = 2\pi k_{B}cT_{V}\nabla_{\mathbf{R}}T_{V}/\hbar\left|\nabla_{\mathbf{R}}T_{V}\right|$
for the relation between the acceleration and vacuum temperature, we obtain
Newton's law of gravity $\mathbf{F}=-GMm\mathbf{R}/R^{3}$ between two
classical objects. In this section, we will consider
the gravitational force when including the quantum wave packet effect in the thermodynamic
origin of gravity.

\begin{figure}
\centering \includegraphics[width=0.8\textwidth]{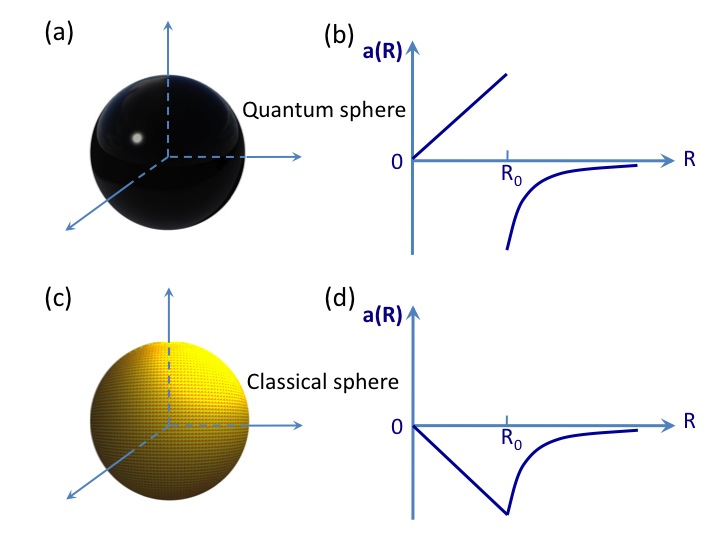} \caption{Fig. (a) shows a wave packet distribution of a particle
in the black sphere. Fig. (b) shows the gravitational acceleration of
this quantum wave packet. Figs. (c) and (d) show the classical situation. \label{fig:quantum sphere}}
\end{figure}

We first consider the following fictional wave function for a fundamental
particle with mass $m_{q}$,
\begin{eqnarray}
\phi_{q}\left(\mathbf{x},t\right) & \simeq & \frac{1}{\sqrt{V}},\left(R<R_{0}\right),\nonumber\\
\phi_{q}\left(\mathbf{x},t\right) & \simeq & 0,\left(R>R_{0}\right).\label{wavefunction}
\end{eqnarray}
 The average density distribution $\left|\phi_{q}\left(\mathbf{x},t\right)\right|^{2}$
is indicated by the black quantum sphere in Fig. \ref{fig:quantum sphere}(a).
Although this is a fictional wave function, its simple form contributes
to our understanding of the abnormal gravitational effect for quantum
wave packets.

\begin{figure}
\centering \includegraphics[width=0.8\textwidth]{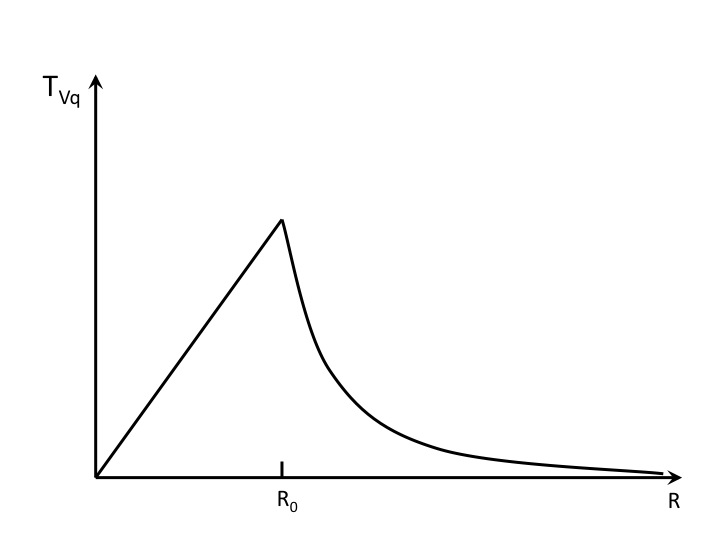} \caption{The vacuum temperature field distribution for a particle with a wave function
given by Eq. (\ref{wavefunction}).\label{fig:QuantumTemperature}}
\end{figure}

For $R > R_{0}$, similar to the consideration of a classical particle,
it is easy to obtain
\begin{equation}
T_{Vq}=\frac{\hbar Gm_{q}}{2\pi k_{B}cR^{2}}.\label{Tq1}
\end{equation}
 At $R = 0$, based on the spherical symmetry, we have ${\bf \mathbf{a}}\left(R = 0\right) = \mathbf{0}$.
Thus, based on the relation between $\bf a$ and $T_{Vq}$, we have $T_{Vq}\left(R=0\right)=0$. Another influence that suggests that $T_{Vq} = 0$ at $R = 0$ is the interference effect of a wave packet
with spherical symmetry. For $R < R_{0}$, again using the spherical
symmetry and Gauss's flux theorem, we have
\begin{equation}
T_{Vq}=\frac{\hbar Gm_{q}R}{2\pi k_{B}cR_{0}^{3}}.\label{Tq2}
\end{equation}
The whole vacuum temperature field distribution is shown in Fig. \ref{fig:QuantumTemperature}. We see that, in the interior of the quantum sphere, the vacuum temperature increases with increasing $R$. This agrees with the physical intuition that the absolute value of the acceleration field increases with increasing $R$ and becomes zero at $R = 0$. In fact, based on the idea presented in the previous section regarding the origin of the gravitational force, the above equation is the only choice for the vacuum temperature field for the quantum sphere shown in Fig. \ref{fig:quantum sphere}(a).

For the quantum wave packet shown in Fig. \ref{fig:quantum sphere}(a), from the relation
between the acceleration field and the vacuum temperature field distribution given
by Eq. (\ref{Unruhvector}), we have
\begin{eqnarray}
\mathbf{a} & = & \frac{Gm_{q}R}{R_{0}^{3}}\mathbf{e}_R,\left(R<R_{0}\right),\nonumber\\
\mathbf{a} & = & -\frac{Gm_{q}}{R^{2}}\mathbf{e}_R,\left(R>R_{0}\right).\label{anomousacc}
\end{eqnarray}
This relation indicates that in the interior of the quantum sphere,
the gravitational force is repulsive. This abnormal gravitational effect is further shown
in Fig. \ref{fig:quantum sphere}(b). It is clear that this abnormal
gravitational effect physically originates from the quantum wave
packet effect for a particle in the black sphere. This phenomenon is reminiscent of the repulsive gravitational effect of dark energy. The application of this repulsive gravitational effect to dark energy \cite{d1,d2,Peeble} and the extension to general relativity will be given elsewhere.

Because the temperature field distribution in Fig. 4 is not continuous for the first-order derivative of $R$ at $R_0$, there is a discontinuity at $R_0$ in the direction of the acceleration field shown in Fig. 3(b) for the gravitational force because we are considering a quantum wave packet. Because, in the calculation of the temperature field, the spatial scale of the phase space is roughly the Planck length, it is expected that the discontinuous region would be of the order of the Planck length. In real-world situations, the quantum wave packet cannot be distributed as shown in Fig. 3(a) in an ideal way. Thus, the discontinuous region will be more complex than the ideal symmetric distribution of the quantum wave packet. The discontinuous distribution of the gravitational field provides a possible way for experimental tests.

If there are $N$ particles in the same quantum state given by Eq.
(\ref{wavefunction}), we have
\begin{eqnarray}
\mathbf{a} & =\frac{2\pi k_{B}c}{\hbar}\sum_{i}\frac{T_{Vi}\left(\mathbf{R}\right)\mathbf{\nabla}_{\mathbf{R}}T_{Vi}\left(\mathbf{R}\right)}{\left\vert \mathbf{\nabla}_{\mathbf{R}}T_{Vi}\left(\mathbf{R}\right)\right\vert }= & \frac{GNm_{q}R}{R_{0}^{3}}\mathbf{e}_R,\left(R<R_{0}\right),\nonumber\\
\mathbf{a} & =\frac{2\pi k_{B}c}{\hbar}\sum_{i}\frac{T_{Vi}\left(\mathbf{R}\right)\mathbf{\nabla}_{\mathbf{R}}T_{Vi}\left(\mathbf{R}\right)}{\left\vert \mathbf{\nabla}_{\mathbf{R}}T_{Vi}\left(\mathbf{R}\right)\right\vert }= & -\frac{GNm_{q}}{R^{2}}\mathbf{e}_R,\left(R>R_{0}\right).\label{anomousacc-1}
\end{eqnarray}

To further understand the abnormal gravitational effect, we consider
a classical sphere with the following density distribution of the mass
\begin{eqnarray}
n\left(\mathbf{x},t\right) & \simeq & \frac{Nm_q}{V},\left(R<R_{0}\right),\nonumber\\
n\left(\mathbf{x},t\right) & \simeq & 0,\left(R>R_{0}\right).\label{densitdis}
\end{eqnarray}
 This density distribution for a classical sphere is shown in Fig.
\ref{fig:quantum sphere}(c). It appears that the same
acceleration field distribution as that of the quantum sphere is found; however, this conclusion is not accurate. For a classical sphere, assuming
there are $N$ particles, the wave packets of
all particles are highly localized. Thus, for a particle at location
${\bf {x}}_{j}$, the vacuum temperature field distribution due to this particle is $T_{Vj}\left({\bf {R}}\right)\sim m_{j}/\left|{\bf {R}-{\bf {x}}_{j}}\right|^{2}$. From ${\bf {a}}\sim\Sigma_{j}T_{Vj}\nabla T_{Vj}/\left|\nabla T_{Vj}\right|$, we get the same result given by Newton's law of gravitation, shown in Fig. \ref{fig:quantum sphere}(d). We stress that the different gravitational force lies in that the quantum states of the quantum sphere and of the classical sphere are different.

Assuming that there are $N$ fundamental particles whose wave functions are $\phi_{1}\left(\mathbf{x},t\right), \cdot\cdot\cdot, \phi_{j}\left(\mathbf{x},t\right), \cdot\cdot\cdot, \phi_{N}\left(\mathbf{x},t\right)$, we can determine the formulas used to calculate the acceleration field due to these $N$ particles. Based on the above special case and the classical limit of Newton's law of gravitation, it is natural to get the following two formulas to calculate the gravitational acceleration of quantum wave packets.
\begin{equation}
\mathbf{a}\left(\mathbf{R},t\right)\mathbf{=}\frac{2\pi k_{B}c}{\hbar}\sum_{j=1}^{N}T_{Vj}\left(\mathbf{R},t\right)\frac{\nabla_{\mathbf{R}}T_{Vj}\left(\mathbf{R},t\right)}{\left\vert \nabla_{\mathbf{R}}T_{Vj}\left(\mathbf{R},t\right)\right\vert },\label{acc}
\end{equation}
 and
\begin{equation}
T_{Vj}\left(\mathbf{R},t\right)=\frac{\hbar Gm_{j}}{2\pi k_{B}c}\left|\int d^{3}\mathbf{x}\phi_{j}^{\ast}\left(\mathbf{x},t\right)\frac{\mathbf{x}-\mathbf{R}}{\left\vert \mathbf{x}-\mathbf{R}\right\vert ^{3}}\phi_{j}\left(\mathbf{x},t\right)\right|.\label{temfield}
\end{equation}
 Here, $m_{j}$ is the mass of the $j$th fundamental particle. The
integral on the right-hand side of Eq. (\ref{temfield}) is due to
the quantum wave packet of the $j$th fundamental particle; the norm of the vector after calculating this integral indicates that $T_{Vj}$ is a scalar field larger than zero. It is easy to
show that if all $N$ particles are highly localized classical
particles, we obtain Newton's law of gravity $\mathbf{a}\left(\mathbf{R}\right) = -\sum_{j}Gm_{j}\left(\mathbf{R-x}_{j}\right)/\left\vert \mathbf{x}_{j}-\mathbf{R}\right\vert ^{3}$,
where $\mathbf{x}_{j}$ is the location of the $j$th particle.

Of course, without full understanding of the quantum gravitational effect, we do not have rigorous derivations of these two equations. In the current situation, it is almost impossible to provide a rigorous derivation (even the Schr\H{o}dinger equation is not obtained with a rigorous derivation). Previous analyses of the thermodynamic origin of gravity
provide important clues and confinement conditions that allow the derivation of these two equations. In a sense, the key
of the present work lies in the determination of the direction of
the gravitational force that gives the correct attractive gravitational effect
for classical objects. The same rule is applied to the case of quantum
wave packets, which predicts a repulsive gravitational effect. In fact,
the absolute value of the acceleration based on Eqs. (\ref{acc})
and (\ref{temfield}) is the same as the result based on Newton's
law of gravity; the only difference lies in the determination of the
direction of the acceleration with the free energy of the whole system
when the thermodynamic origin of gravity is considered.


\section{Experimental scheme to test the repulsive gravitational effect}

\begin{figure}
\centering \includegraphics[width=0.8\textwidth]{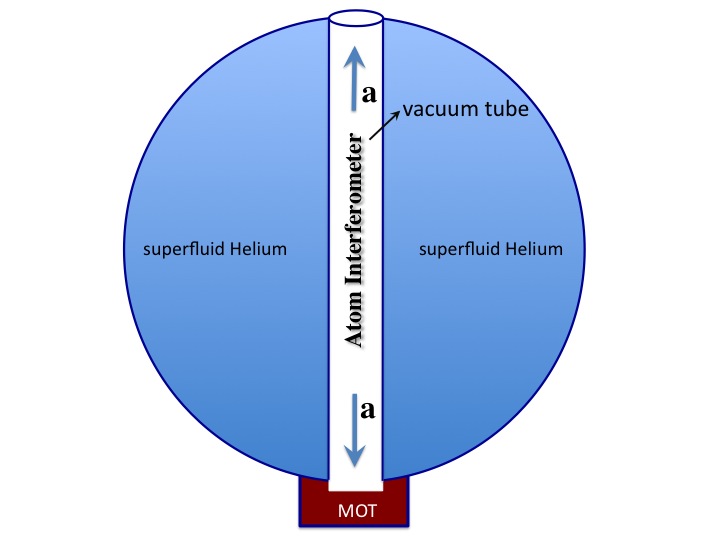} \caption{An experimental scheme to test the abnormal
gravitational effect. Various
apparatuses measuring the acceleration of gravity are placed in the hole in the superfluid helium sphere. As an example,
we consider the application of an atom interferometer, where the vacuum
tube of the interference region is placed in the interior of the superfluid
helium sphere while the magneto-optical trap (MOT) of the
cooling and trapping cold atoms may be placed outside the sphere.\label{fig:ExpScheme}}
\end{figure}

Now, we turn to consider an experimental scheme to test the
repulsive gravitational effect using superfluid $^{4}$He, as shown in Fig. \ref{fig:ExpScheme}.
For the sake of simplicity, we consider a sphere full of superfluid $^{4}$He.
There is a hole in this sphere. In this situation, from Eqs. (\ref{acc})
and (\ref{temfield}), the gravity acceleration in the hole of the superfluid $^{4}$He sphere can be approximated as
\begin{equation}
\mathbf{a} = \frac{4\pi}{3}Gn_{He}\mathbf{R}.\label{ahe}
\end{equation}
Here, the liquid helium's density is $n_{He} \approx 550$ kg/m$^{3}$.
From this equation, the anomalous acceleration is $\mathbf{a} = 1.5\times10^{-7}\mathbf{R}/s^{2}$.
The gradient of this anomalous acceleration is $1.5 \times 10^{-7}/s^{2}$.
Even if only the condensate component of the superfluid $^{4}$He is considered,
for a sphere of radius $1m$, the maximum
anomalous acceleration due to the condensate component is roughly $10^{-8} m/s^{2}$.
It is noteworthy that this value is well within the range of the presently available experimental
technique of atom interferometry \cite{AI}, which is used to measure the acceleration of gravity. Nevertheless, this is a very weak observable effect.
Thus, it is unlikely that we will find evidence to verify or falsify this
anomalous acceleration without future experiments. Apart from atom
interferometry, the measurement of acceleration due to gravity using Bloch
oscillation \cite{Bloch} for cold atoms in optical lattices, superfluid
helium interferometry \cite{HeliumAI}, free-fall absolute gravimeters
\cite{Freefalling}, and superconducting gravimeters \cite{superconductor,Paik}
provides other methods to test the abnormal gravitational effect.
In particular, the standard deviation of free-fall absolute gravimeters
in the present technique is roughly $10^{-8}m/s^{2}$; superconducting
gravimeters have achieved sensitivities of one thousandth of one
billionth ($10^{-12}$) of the Earth's surface gravity.

Because this abnormal gravitational effect could be tested using contrast
experiments with superfluid helium and normal helium, and based on the idea that the abnormal gravitational effect is location-dependent, the
sensitivities of the gravimeter could be used to test the abnormal
gravitational effect. For this reason, it is very promising to test the abnormal gravitational effect in future experiments.

One possible obstacle to a definitive test of the abnormal quantum gravity effect with a superfluid helium sphere is found in our current understanding of
the superfluid behavior of liquid helium. In the ordinary understanding
of superfluid helium, the superfluid fraction can reach nearly $100\%$
while the condensate fraction is roughly $8\%$. Because of the strong
interaction between helium atoms, the liquid helium is a very complex,
strongly correlated system. As there are many open questions regarding strongly correlated systems, we cannot exclude the possibility that the wave packets of all helium atoms
are localized, although the whole system still exhibits superfluid
behavior. This significantly differs from the Bose-Einstein condensate
in dilute gases, in which the wave function of the atoms in the condensate
is delocalized throughout the condensate. If the wave packets of all of the helium atoms are localized, we cannot observe the repulsive gravitational effect.


\section{Summary and discussion}

In summary, using the thermodynamic origin of gravity, we consider the gravitational effect for classical and quantum objects. We found evidence for a repulsive gravitational effect for quantum wave packets, which can be tested using currently available experimental techniques.
Starting from the present phenomenological studies of the gravitational effects of quantum wave packets,
it would be useful to consider whether this macroscopic,
abnormal gravitational effect could be derived from the microscopic
mechanisms of quantum gravity such as superstring with a positive
cosmological constant \cite{Kachru}, loop quantum gravity and twistor
theory \cite{Penrose}, etc. Recently, the relation between
gravity and thermodynamics has been studied within the framework of loop quantum gravity \cite{Smolin}, which provides a possible solution to this problem.

Although the microscopic interaction mechanisms of quantum gravity
at the Planck scale are an unsolved problem, the thermodynamics of the
macroscopic quantum gravitational effect presented in this work are still meaningful.
When statistical mechanics was developed by Boltzmann in 1870, the
concept of the atom was still an unrecognized hypothesis (not to mention
the collision mechanisms between atoms due to electromagnetic interaction).
However, this does not influence the power of statistical mechanics
for the description of gas dynamics. On the other hand, the theoretical
and experimental advances of statistical mechanics greatly promoted
the further understanding of atoms. If the thermodynamic origin of gravity can be verified by future experiments and astronomical observations, investigation into the
abnormal gravitational effect will also promote the understanding
of quantum gravity at the Planck scale.

\section*{Acknowledgements}
We thank the discussions with Prof. Biao Wu, and his great encouragements. We also thank the great encouragements of Prof. Frank Wilczek, Betsy Devine, and Prof. W. Vincent Liu. This work was supported by National Key Basic Research and Development Program of China under Grant No. 2011CB921503 and NSFC 11175246, 11334001.

\end{document}